\documentstyle[11pt,newpasp,twoside,epsfig]{article}
\def\gtsim{\lower.5ex\hbox{$\; \buildrel > \over \sim \;$}}
\def\ltsim{\lower.5ex\hbox{$\; \buildrel < \over \sim \;$}}

\begin{document}

\title{Excitation of H$_2$ and HD in Shocks and PDRs}
\author{
  F.\ Bertoldi$^1$,
  B.\ T.\ Draine$^2$,
  D.\ Rosenthal$^3$,
  R.\ Timmermann$^3$, 
  S.\ K.\ Ramsay Howat$^4$, T.\ Geballe$^5$,
  H.\ Feuchtgruber$^3$, and S.\ Drapatz$^3$
} 
\affil{
  $^1$ Max-Planck-Institut f\"ur Radioastronomie, D-53121 Bonn, Germany\\
$^2$ Princeton University Observatory, Princeton, NJ 08540, USA\\
$^3$ Max-Planck-Institut f\"ur extraterrestrische Physik, 
       D-85740 Garching, Germany\\
  $^4$ Royal Observatory, Blackford Hill, Edinburgh, EH9 3HJ, UK\\
  $^5$ Gemini Observatory, Hilo, HI 96720, USA
}

\begin{abstract}
  Photodissociation regions (PDRs) and shocks give rise to conspicuous
  emission from rotationally and vibrationally excited molecular
  hydrogen. This line emission has now been studied with ISO and from
  the ground in great detail. A remarkable discovery has been that
  toward the Orion outflow and other shock-excited regions, the H$_2$
  level populations show a very high excitation component.  We suggest
  that these high-excitation populations may arise from non-thermal
  pumping processes, such as H$_2$ formation and high-velocity
  ion-molecule collision in partially dissociative shocks.  In PDRs such
  as NGC\,7023 however, formation pumping is always less important than
  fluorescent
  pumping.

  We furthermore present two HD emission line detections toward Orion
  Peak 1. This enables the first comparison of the H$_2$ and the
  HD excitation, which surprisingly turn out to be identical.
\end{abstract}



\section{Introduction}

Shocks and photodissociation regions show strong H$_2$ emission which is
detectable in the visible, near- and mid-infrared.  Among the important
H$_2$ lines observed are those of pure rotational transitions in the
vibrational ground state, at wavelengths between about 2.4 and 28
$\mu$m.  This spectral range has been difficult to access from the
ground, and only recently has the {\it Infrared Space Observatory} (ISO)
opened up this window to sensitive spectroscopy. Much attention has
focused on the pure rotational emission, because it provides important
clues about the physical state of the emitting gas.  ISO and
ground-based observations were able to trace the populations of levels
at high energy, up to near the dissociation limit at 4.5 eV above the
ground state.  Excitation diagrams such as those shown in
Figs.~1 or 4 indicate that in shocks and
PDRs (see Wright, this volume), the higher energy levels typically show
remarkably high excitation temperatures, of order 2000 to 3000~K.  Could
this emission from high levels actually be thermally excited\,?  Although
in principle possible, the kinetic temperatures and gas densities
required for their thermalization seem uncomfortably high. What then
could be the cause of this high excitation\,? In the following, we will
briefly discuss several possible mechanisms.

\section{H$_2$ in Shocks: the Orion Outflow}

OMC-1, the molecular cloud behind the Orion M42 Nebula, is the
best-studied massive star-forming region.  It embeds a young stellar
object which produces a spectacular outflow that shocks the surrounding
molecular gas, producing strong H$_2$ infrared line emission fragmented
in a multitude of clumps and arcs (Schultz et al.\ 1998).  Peak~1
(Beckwith et al.\ 1978) is the brighter of the two H$_2$ emission
concentrations of the outflow.  Toward Peak 1 we obtained 2.4--45
$\mu$m spectra with ISO, which show over 60 $\rm H_2$ ro-vibrational and
pure rotational lines, the latter ranging from 0--0 S(1) to 0--0 S(25)
(Rosenthal et al.\ 1999,\,2000).  The extinction-corrected level column
density distribution derived from the observed line intensities is shown
in Fig.~1 ({\it left}).

\begin{figure}
\centering
\epsfig{file=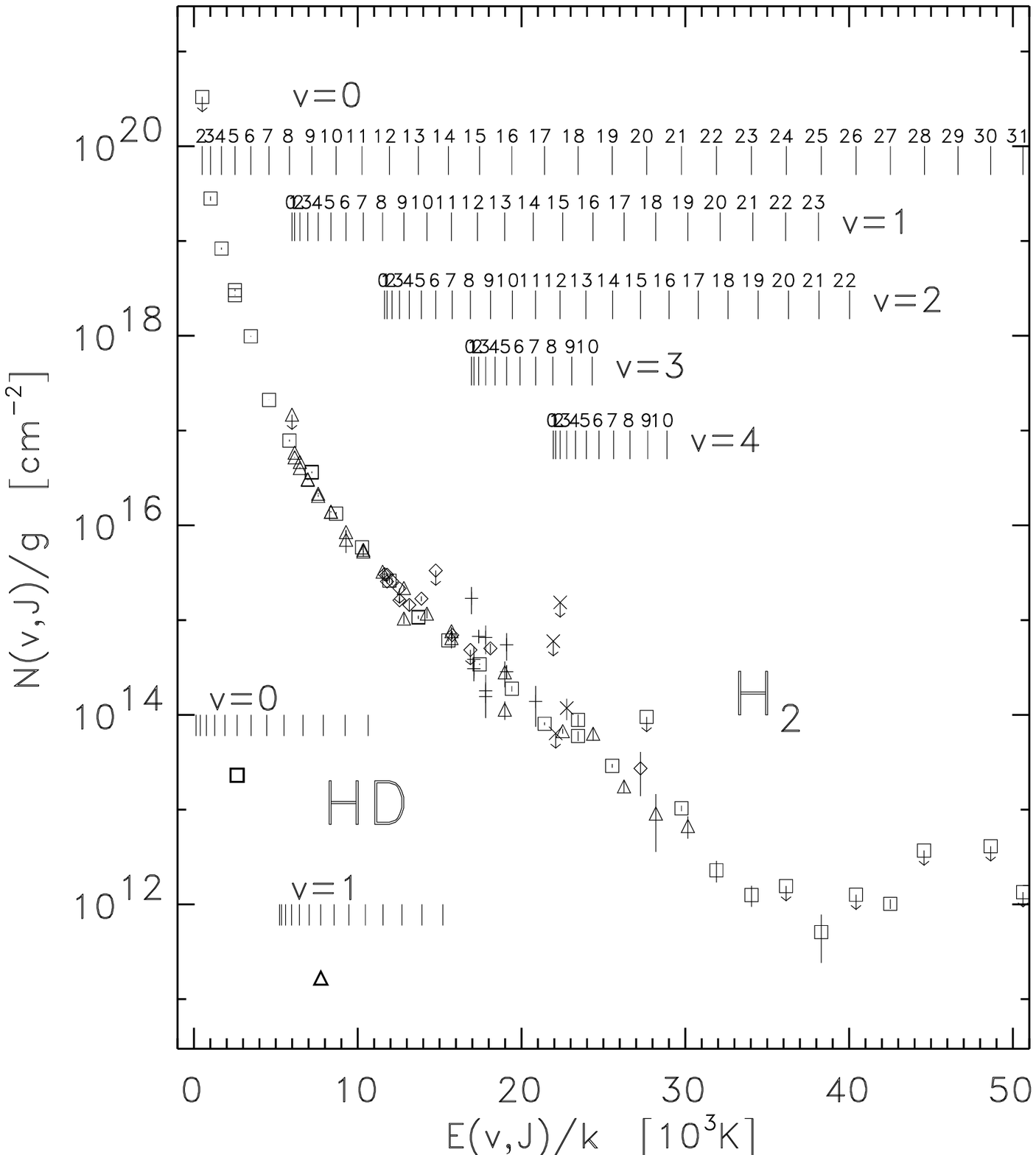,width=0.49\textwidth} 
\epsfig{file=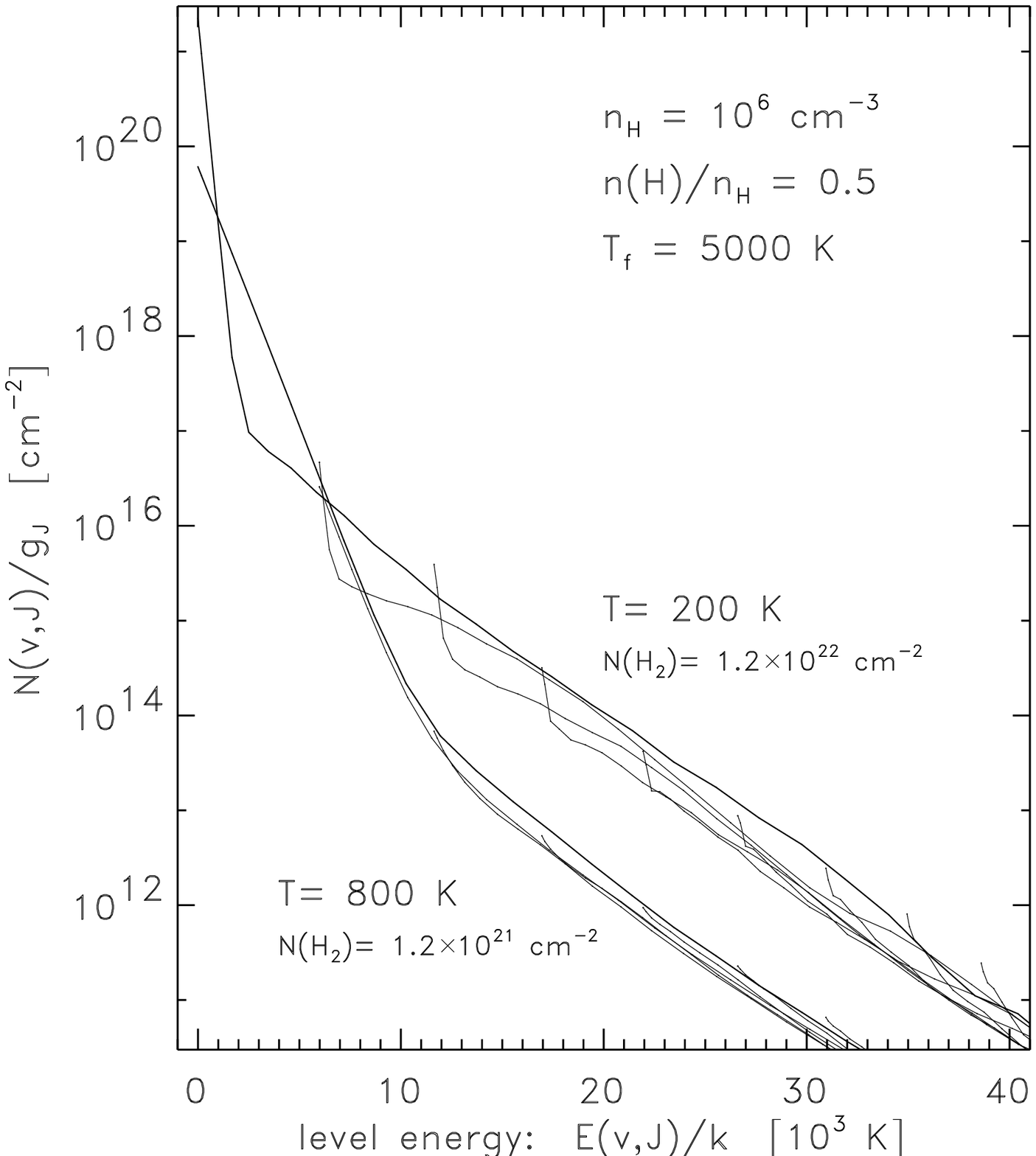,width=0.49\textwidth} 
\caption{{\it Left:} Extinction-corrected, observed H$_2$ and 
  HD level column densities toward Orion Peak 1, divided by their
  degeneracy, plotted against the level energy $E(v,J)$.  {\it
    Right:} Level populations predicted by non-LTE models of two gas
  layers with the labeled temperatures and column densities.  Individual
  vibrational level distributions are shown as separate lines.  The sum
  of the two distributions well matches the observed level
  distribution toward Peak 1 shown in the left panel.}
\label{fig:boltzmann}
\end{figure}

\subsection{Excitation of high-energy H$_2$ levels toward Peak 1} 

Only a fraction $\sim$\,10$^{-3}$ of the total column density 
1.9\,$\times 10^{21}$ cm$^{-2}$ of warm H$_2$ toward Peak 1 resides in high energy
levels, $E(v,J)/k$\,$>$\,10,000~K.  For these levels the distribution of
column densities implies an excitation temperature, $T_{ex}$, of about
3000 K; $T_{ex}$ is defined by a fit to the column densities of the form
$N(v,J)/g$\,$\propto$\,$\exp(-E(v,J)/kT_{ex})$.  If the observed $N(v,J)$ is
decomposed as a sum of LTE distributions with a range of excitation
temperatures, then we find that about 99\% of all warm H$_2$ has
$T_{ex}$\,$<$\,2000 K, and 90\% has $T_{ex}$\,$<$\,1000 K (see Rosenthal et al.\
2000).  If the distribution of excitation temperatures were to reflect
the actual distribution of kinetic temperatures, such a steep
distribution of temperatures would be difficult to reconcile with the
expected smooth temperature profile of planar C-type shocks.  In C-type
shocks the gas temperature changes smoothly, and a large fraction of the
warm gas is near the {\it maximum} temperature reached in the shock
(Timmermann 1996).  Even with a distribution of shocks of different
velocities and a corresponding range of peak temperatures, an excitation
temperature distribution like that found toward Peak 1 is difficult to
construct, unless one were to combine shocks with very disparate filling
factors, e.g., several percent at 40 km s$^{-1}$, the rest at 20 km
s$^{-1}$. Even in bow shocks, which are often cited to match the
observed excitation (Smith 1991), the velocity changes slowly with
distance from the apex, and the observed distribution of temperatures
would hardly be expected.

In dissociative J-type shocks (e.g.\ Hollenbach \& McKee 1989), 
the molecules are destroyed in the shock,
and they reform in a postshock layer where the temperature has dropped
much below 3000 K, somewhat dependent on the H$_2$ formation rate at
higher temperatures, which is very uncertain (e.g.\ Bertoldi 1997).  
Dissociative J-shocks can
therefore also not account for the high excitation H$_2$ we observe.

Even if temperatures of 3000 K or more can be reached in nondissociative
shocks, the higher H$_2$ levels would remain sub-thermally excited unless
the gas density is high enough that the collisional excitation and
deexcitation rates are comparable to those for radiative decay. A
``critical'' gas density can be estimated for a given level as that
density for
which the total collisional deexcitation rate of the level equals its
total radiative decay rate.  We find that even at kinetic temperatures
of 3000 K, gas densities well above $10^6$ cm$^{-3}$ would be needed
to maintain the higher $v$=0 levels at populations resembling LTE.
Since such high densities may not prevail in the shocked gas of the
Orion outflow, we explore mechanisms other than thermal excitation
that might account for high populations in the upper energy states.

\subsection{Time-dependent C-shocks}

When a high velocity outflow strikes dense molecular gas, and a C-type
shock is first established, J-type shocks can temporarily form within
the C-shock (Chi\`{e}ze et al.\ 1998; Flower \& Pineau des For\^ets
1999).  In such an embedded J-shock, a small column of molecular gas is
briefly heated to high temperatures without being dissociated, and if
the gas density is sufficiently high, this would produce a
high-excitation tail in the column density distribution.  The lifetime
of the embedded J-shock is small, and the high-excitation populations
are transient, unless shocks were continuously reforming.  Embedded
J-shocks may also form when a C-shock encounters a dense clump.  Since
the time scale for the fading of the high-excitation tail in newly
forming C-shocks is somewhat smaller than the age of the Orion outflow,
we should expect spatial variations in the upper level excitation.
Whether such gradients exist is unknown, but it would provide an
interesting test for the transient J-shock model.

\subsection{H$_2$ formation pumping}
 
H$_2$ formation is one other mechanism to populate the high-energy H$_2$
levels.  Molecular hydrogen is believed to form on the surfaces of dust
grains.  Some fraction of the 4.5~eV released during the formation of an
H$_2$ molecule is required to leave the grain, and the remainder is
split between translation, rotation, and vibration of the new molecule.

Toward Peak 1 we found that a fraction 6.8\,$\times 10^{-4}$ of the total
warm H$_2$ column is in states with energy $E(v,J)/k$\,$\geq$\,10,000 K.
Could H$_2$ formation in a steady state account for such a fraction of
molecules in highly excited states\,?  We can try a very simple estimate.
The pumping rate due to formation pumping is equal to the H$_2$
formation rate, $n({\rm H})\, n_{\rm H}\, R_{gr}$, where we adopt
$R_{gr}$\,$\approx$\,5\,$\times 10^{-17}$ cm$^3$ s$^{-1}$ as the H$_2$
formation rate coefficient per hydrogen nucleus. The radiative decay
rate we estimate from a characteristic radiative lifetime of a level at
$\sim$\,2 eV, which is $\sim$\,10$^{6}$ sec, 
and we prolong this by an average number of 
$\sim$\,5 transitions for the molecule 
to reach its ground state. Thereby defining an effective
$A$-coefficient for the decay rate of highly excited states, 
$A_x$\,$\approx$\,2\,$\times 10^{-7}$ s$^{-1}$, the population balance for the excited
states writes
\begin{equation}
  R_{gr}\, n_{\rm H}~ n({\rm H}) = n_x({\rm H_2})\, A_x, 
\end{equation}
which yields an excited H$_2$ fraction
\begin{equation}
 {n_x({\rm H_2})\over n({\rm H_2}) } 
= {n{(\rm H)}\over n({\rm H_2})} {n_{\rm H} R_{gr} \over A_x }
= 5\times 10^{-4} \left(n_{\rm H}\over 10^6\,{\rm cm^{-3}}\right)
\left(n{(\rm H)}\over 2\,n({\rm H_2})\right)~,
\end{equation}
that agrees well with the observed value.  This simple estimate shows
that H$_2$ formation could in principle
account for some of the high-excitation level
populations, provided the density is high enough, 
the atomic fraction is not small, and
the formation rate coefficient in the warm shocked gas is comparable
to the value implied at $\approx 100$~K from Copernicus
observations, $R_{gr}$\,$\approx$\,3\,$\times 10^{-17}$~cm$^3$~s$^{-1}$ (Jura
1975).

Can H$_2$ formation also 
produce the observed shape of the high energy level populations\,? 

The energy required to desorb from the dust grain depends on whether the
atoms are bound by van der Waals forces (physisorbed, $\sim$\,0.03 eV),
such as on water ice, or by chemical bonds (chemisorbed, 
$\sim$\,2 eV), like on a silicate surface.
The exact level distribution of newly formed H$_2$ is unknown, but
it could very well contribute to the observed excitation at intermediate
energies, $E(v,J)$\,$\approx$\,1--3~eV.  Experiments and theoretical calculations
are yet inconclusive regarding the excitation of a newly
formed, desorbed hydrogen molecule.  Hunter \& Watson (1978) consider
physisorption and find a partition between rotational, vibrational, and
translational energy, $E_r$\,$\approx$\,0.8~eV, $E_v$\,$\approx$\,3.3~eV,
$E_t$\,$\approx$\,0.8~eV.  Duley \& Williams (1986) assume the hydrogen to be
physisorbed on silicate surfaces and find $E_r$\,$<$\,0.01~eV, 
$E_v$\,$\approx$\,3.3~eV, $E_t$\,$\approx$\,0.2~eV.  Duley \& Williams (1993) find that on PAH
and graphite surfaces, $E_r$\,$<$\,0.01~eV, $E_v$\,$<$\,10$^{-3}$~eV, whereas Duley
\& Williams (1993) and Leonas \& Pjarnpuu (1981) find that on H$_2$O ice
and polymeric hydrocarbon surfaces, $E_v$\,$\approx$\,4~eV.  Evidently, there
is no conclusive agreement yet, except for a
trend to expect that newly formed molecules must be
vibrationally hot and rotationally cold.

Using the PDR codes of Draine \& Bertoldi (1996),
we computed the non-LTE level populations of partially dissociated
gas of fixed density and temperature, 
and assuming that the H$_2$ formation distribution follows
\begin{equation}
\phi(v,J) = \phi(0,0)~ g_n(J)~ g_f ~ e^{-E(v,J)/k T_{f}}~,
\end{equation}
where the nuclear degeneracy factor, $g_n(J)$=1 for even $J$, 3 for
odd $J$. The formation temperature, $T_f$, is a free parameter, and so
is the formation degeneracy, $g_f$. 
The exponential form of the level distribution (Black \& van Dishoeck
1987) is somewhat arbitrary,
but in lack of observational constraints, seems reasonable.
Fig.~2 shows the computed
level populations. The high excitation tail in the 
population of levels with energy $>$\,5000 K is due to formation pumping,
whereas the lower levels are populated through collisional excitation.
The difference in the level population when adopting
$g_f$=2$J$+1 or $g_f$=1+$v$
is illustrated in Fig.~2 ({\it left}). 
When emphasizing vibrational 
excitation, the individual vibrational level populations
show a splitting similar to that seen for fluorescent excitation.
When the weight lies on rotational excitation, the level
populations remain tighter, more like what we observe
toward Peak 1.
Fig.~2 ({\it right}) 
shows how the level populations depend on the 
value of the formation temperature, $T_f$, where we adopt
$g_f$=2$J$+1.

No single temperature gas slab is able to reproduce the level
distributions observed toward Peak 1. However, the sum of two components
with hydrogen nuclei density $n_{\rm H}$\,=\,10$^6$ cm$^{-3}$, atomic
fraction $n({\rm H})/n_{\rm H}$\,=\,0.5, and respective 
temperatures of 200~K and 800~K, 
with column densities 
$N_{\rm H_2}$\,=\,1.2\,$\times 10^{22}$ cm$^{-2}$ and 
1.2\,$\times 10^{21}$ cm$^{-2}$, can reproduce the observed
level column distribution very well (Fig.\ 1, {\it right}).  
In order
to produce a narrow distribution of level columns in the Boltzmann plot,
we would have to assume pronounced rotational excitation
of the newly-formed H$_2$.
This is somewhat contrary to the expectation of vibrationally
hot, but rotationally cold new molecules.

\begin{figure}
\centering
\epsfig{file=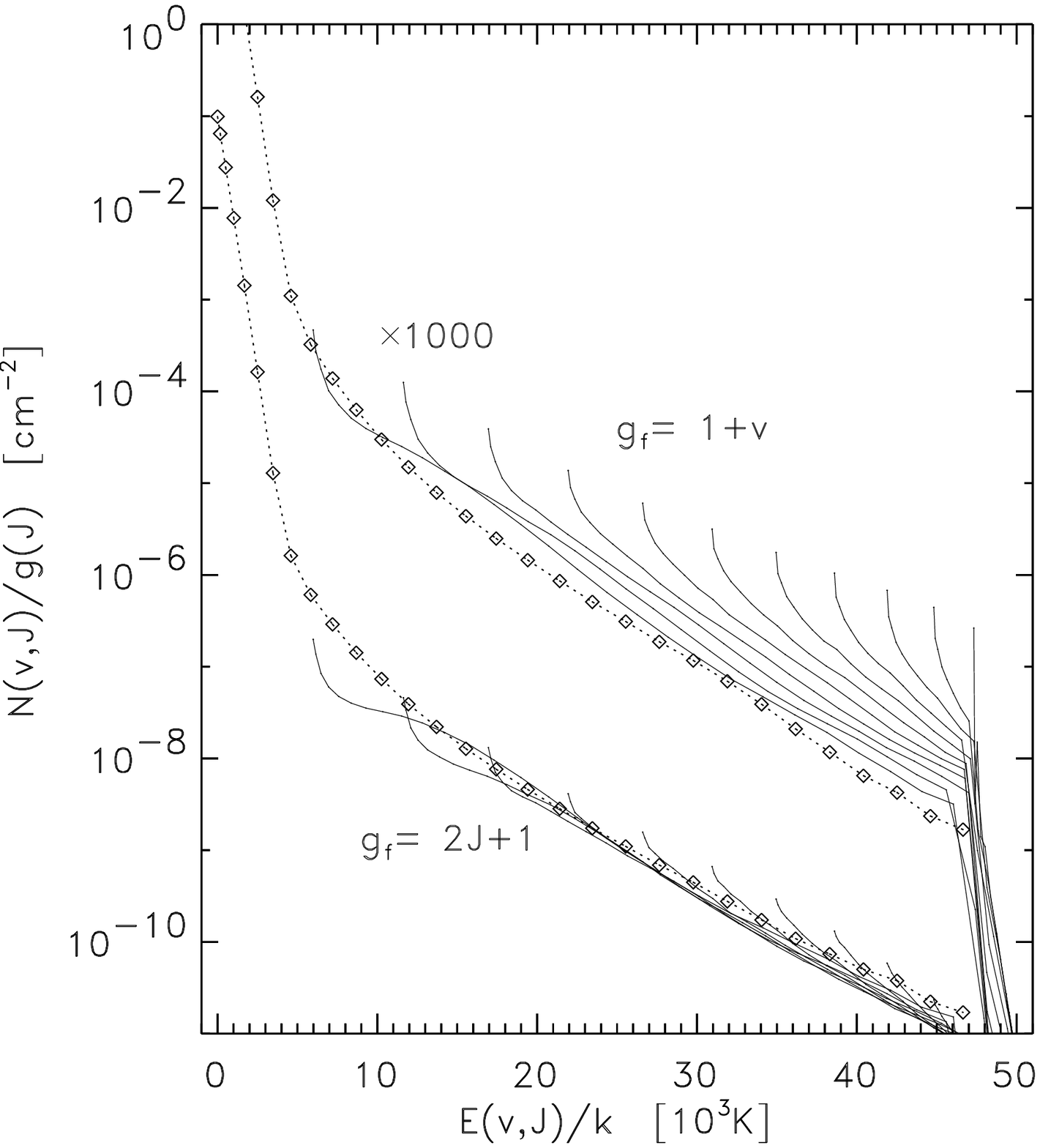,width=0.48\textwidth} 
\epsfig{file=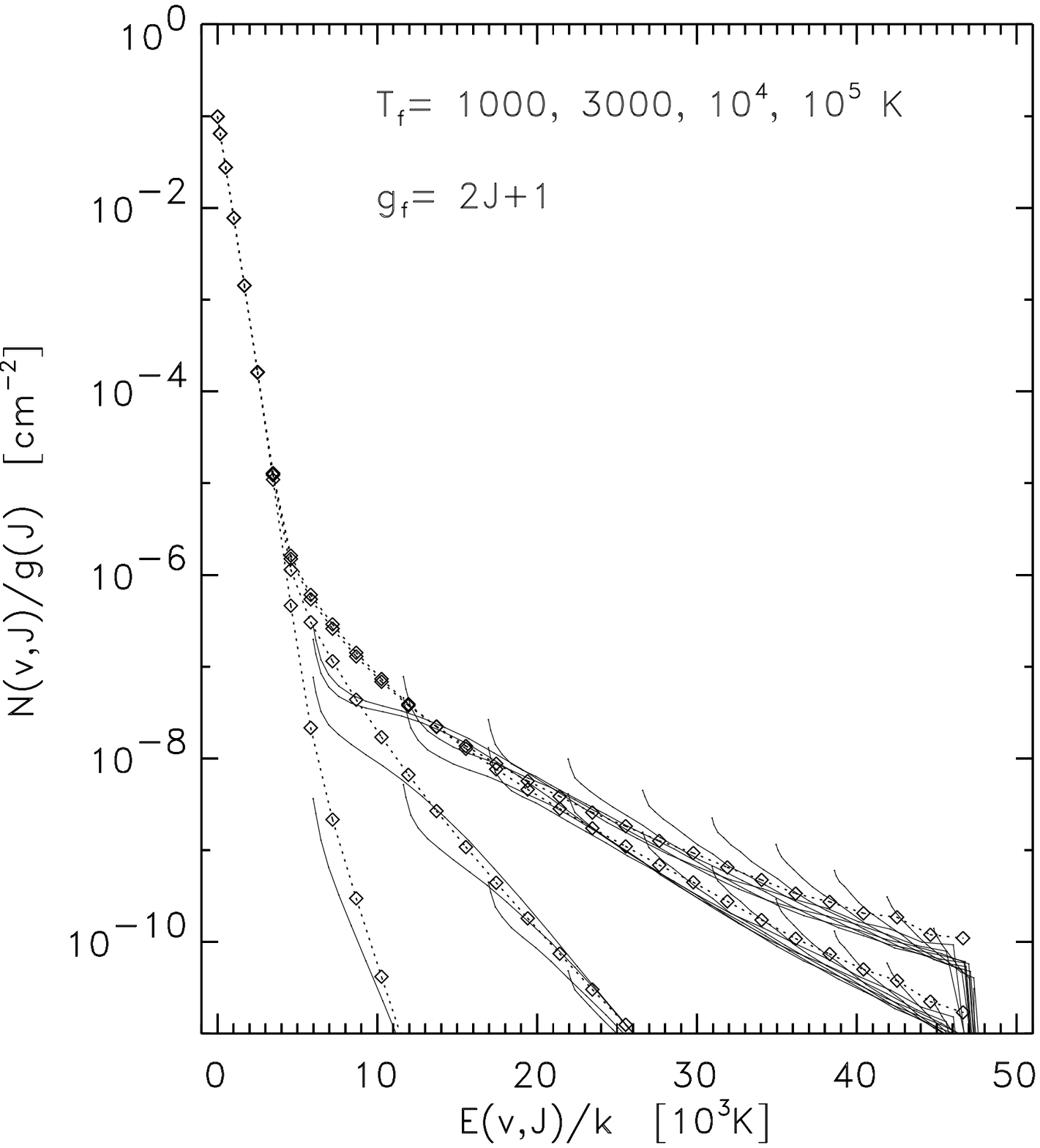,width=0.48\textwidth} 
\vspace{-.2cm}
\caption{Level populations for non-LTE models 
  including collisional processes and H$_2$ formation, assuming a gas
  temperature of 400 K, a number density of H nuclei of 
$10^5$ cm$^{-3}$, and an atomic fraction of 0.5, i.e.,
  $n({\rm H})$\,=\,2\,$n({\rm H_2})$. {\it Left:} Two different
  distributions are assumed for newly-formed H$_2$, one prefering
  vibrational, the other rotational excitation. {\it Right:} Comparison
  of level populations resulting from different H$_2$ formation
  temperatures.}
\label{fig:excform}
\end{figure}

\subsection{Non-thermal collisions}
 
An even more important pumping mechanism for the high-excitation levels
could be non-thermal collisions between molecules and ions in a magnetic
shock.  In magnetic C-type shocks, which are believed to be responsible
for most of the emission in Peak~1, the gas is accelerated through fast
inelastic collisions. In a magnetic precursor the ions, which are tied
to the magnetic field, collide with the undisturbed preshock gas at
relative velocities comparable to the shock speed. Such non-thermal
ion-molecule collisions lead to the acceleration of the molecules and to
their internal excitation. High-velocity molecules subsequently collide
with other molecules, resulting in a cascade of collisions during which
the relative kinetic energy is in part converted to internal excitation
of the molecules (O'Brien \& Drury 1996).  In sufficiently fast
C-shocks, the ion--H$_2$ and H$_2$--H$_2$ collisions can lead to a
significant collisional dissociation rate.  The molecules dissociated in
a partially dissociative, steady-state shock reform further downstream,
so that across such a shock the H$_2$ dissociation rate equals the H$_2$
reformation rate. For every collisionally dissociated molecule there
is likely to be a much 
larger number of inelastic collisions which did not lead to
dissociation, but to the excitation of the molecule into high
rovibrational states.  The
high-excitation H$_2$ level column densities thereby created should
then be larger than those caused by H$_2$ formation alone.

We conclude that non-thermal collisions in partially dissociative
shocks could pump the high-excitation states in the H$_2$ electronic
ground state to the levels observed. No detailed shock models are
available yet which account for this process.

\begin{figure}[h]
\label{fig:00S25}
  \begin{minipage}[t]{0.49\linewidth}
  \epsfig{file=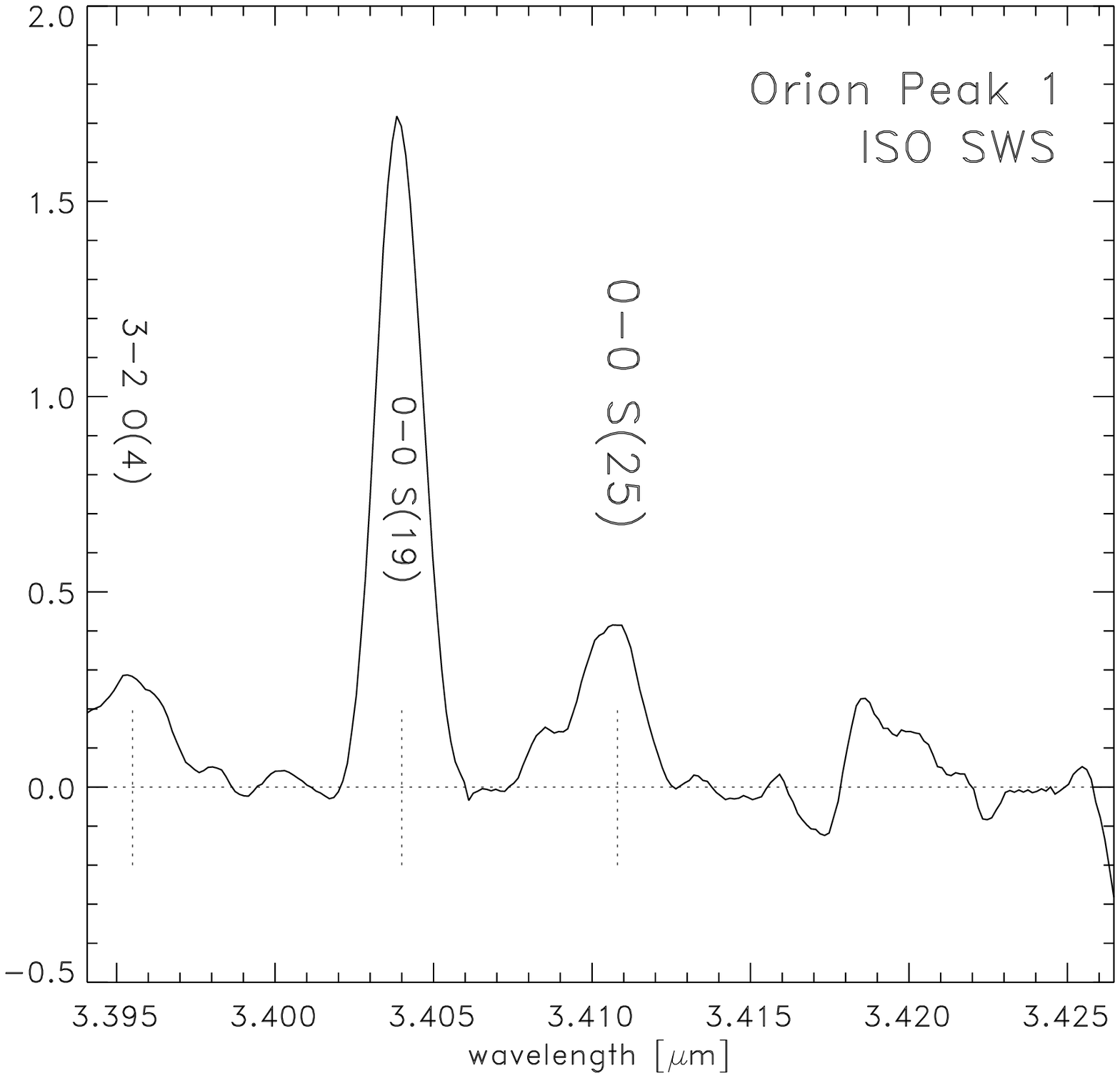,width=\textwidth} 
  \end{minipage}
  \begin{minipage}[t]{0.49\linewidth}
  \vspace*{-6.cm}
 \caption{ H$_2$ $v$=0--0 S(25) observed with the ISO SWS toward
   Orion Peak 1. The line appears clearly in this spectrum taken in the
   AOT 02, $\sim$\,0.01$\lambda$ range grating scan mode, and also appears in
   the shallower AOT 01 full (2.4--45\,$\mu$m) scan not shown here (see
   Rosenthal et al.\ 2000).  }
  \end{minipage}
\end{figure}

\subsection{0--0 S(25)}

The $J$=27 level observed through the 0--0~S(25) line
(Fig.~3) appears overpopulated by a factor of seven (see
Fig.~1) over what would be expected from an
extrapolation of the lower energy level populations. The $J$=27 level
lies 3.6~eV above ground, only 0.9~eV from the dissociation limit. H$_2$
that is newly formed on grains is less likely to populate rotational
states so high, as some fraction of the formation energy is lost to
overcome the grain surface potential, and some goes into vibrational
excitation and translational kinetic energy.

Unless we misidentified the 0--0~S(25) line, it appears that a different
mechanism may be populating this level, and possibly other high levels.
The gas phase formation of H$_2$ via H$^-$, e.g., might be able to leave
the new molecule in such a high rotational state.

\section{Excitation of H$_2$ in PDRs}

A ``photodissociation region'' is the interface separating a region
which is predominantly molecular, and a region where the ultraviolet
energy density is sufficiently high that hydrogen is mostly atomic.
Young O and B stars located near molecular clouds produce conspicuous
PDRs.  A number of general reviews of PDRs have appeared recently
(Hollenbach \& Tielens 1997,\,1999; Sternberg, Yan, \& Dalgarno 1998;
Walmsley 1998), and the current status of PDR modeling with regard to
ISO observations was discussed by Draine \& Bertoldi (1999).

The structure of a PDR is determined primarily by the attenuation of
the far-ultraviolet (6--13.6 eV) radiation field, as one moves from the
ionization front into the PDR.  The dominant process is the
photodissociation of H$_2$, the rate for which is determined by both
H$_2$ self-shielding (Draine \& Bertoldi 1996) and  attenuation by
dust.

The $(v,J)$ ro-vibrational excited states of H$_2$ are populated by
inelastic collisions, by UV pumping, and by formation on grains.  At the
typical densities $n_{\rm H}$\,$\gtsim$\,10$^4$ cm$^{-3}$ of bright PDRs,
collisions maintain the ($v$=0,~$J$) levels of H$_2$ in approximate
thermal equilibrium for $J$\,$\ltsim$\,9.  Therefore measurements of the
quadrupole emission line intensities with ISO provide a good indicator
of the gas temperature.  ISO observations toward the PDRs in, e.g., S140
(Timmermann et al.\ 1996), NGC\,7023 (Fuente et al.\ 1999), or NGC\,2023
(Draine \& Bertoldi 2000), provided unequivocal evidence for gas
temperatures in the 500--1000 K range in a portion of the
photodissociation region (PDR) where the H$_2$ fraction is appreciable.
These temperatures were higher than expected from current models of the
heating and cooling processes in PDRs, and therefore require
reconsideration of the physics of the gas and dust in PDRs.

The higher energy H$_2$ levels are not populated by collisions, but by
the decay of electronically excited states, which were pumped through
the absorption of FUV photons. Because most molecules are excited from
their ro-vibrational ground states, most subsequent transitions change
the rotational quantum number, $J$, only by one or two, and the decay back
into the electronic ground state populates states within a limited range
of $J$.
The cascade down the vibrational states to the ground states can
then significantly populate states up $J$=10 or beyond. 
Because $J$=0,\,1 pose
a reflective boundary for the cascade down, 
the lowest $J$ levels within each vibrational
state build up enhanced populations, which results in the characteristic
population distribution of fluorescently excited H$_2$ displayed in
Fig.~4.

\begin{figure}[tb]
\centering
\epsfig{file=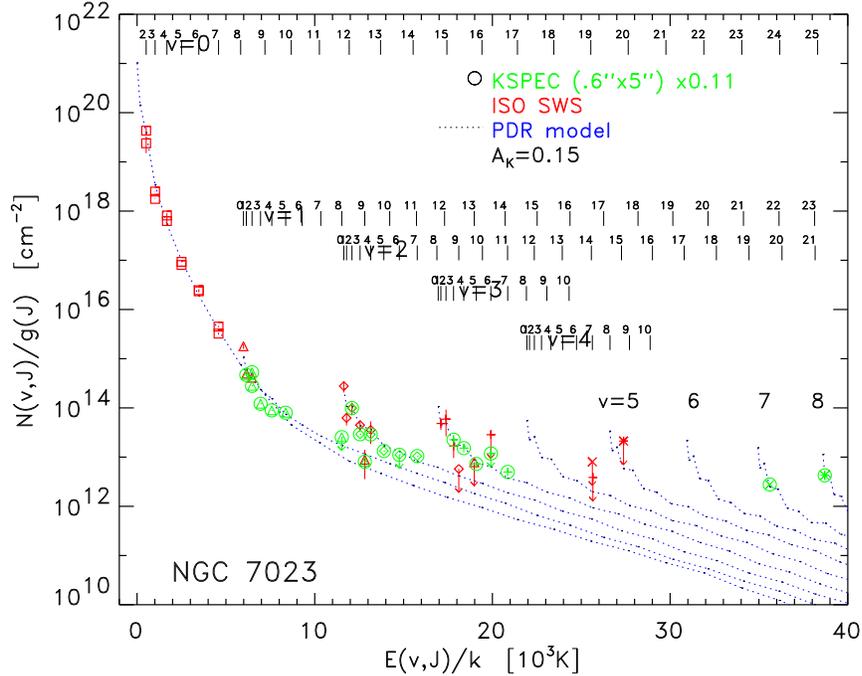,width=0.7\textwidth,angle=90} 
\vspace{-.2cm}
\caption{ Extinction-corrected H$_2$ level populations
  observed toward NGC\,7023, plotted against the level energy.  ISO and
  ground-based near-IR observations (Martini et al.\ 1997; circled
  symbols), scaled to match the ISO populations, are shown.  For $v$=0,
  $J$=2--7 two ISO data points are shown: the respective upper ones
  are adopted from Fuente et al.\ (1999).  Our PDR model for the level
  populations is traced by dotted lines.}
\label{fig:7023}
\end{figure}

As an example we compare the level populations observed toward a bright
filament in the NGC\,7023 PDR with our best fit PDR model, which adopts
$n_{\rm H}$\,=\,10$^5$ cm$^{-3}$ and an incident FUV flux
$\chi$\,=\,5000\,times\,1.2\,$\times 10^7$ cm$^{-2}$ s$^{-1}$. 
The observations shown include our ISO as well as ground-based
spectroscopy. Our PDR model includes all basic collisional
and radiative processes, including detailed thermal balance (Draine \&
Bertoldi 1996,\,1999,\,2000). The agreement between the model and the
observations is remarkably good considering
the simplicity of a constant density, planar slab model.

Black \& van Dishoeck (1987), Le Bourlot et al.\ (1995), and 
Bertoldi (1997) questioned whether processes other than collisions and
fluorescence could affect the high energy or high rotational state
populations in PDRs.  Burton et al.\ (1992) suggested that H$_2$
formation might be responsible for an apparent excess in the $v$=4
levels observed in the NGC\,2023 PDR.

Would we expect formation pumping to have a significant effect on the
high-excitation level populations in PDRs\,? Of all photo-excitations of
H$_2$ by FUV photons, about 10 to 15\% lead to dissociation, the
remaining to pumping of high-excitation levels in the electronic ground
state. In equilibrium, dissociation is balanced by reformation of H$_2$,
so that the ratio of the rates of formation pumping and fluorescent
pumping of the high-excitation levels in the electronic ground state is
$\sim$\,15/85. Overall, fluorescent pumping should therefore dominate
over formation pumping by a factor five.  Unless the level distribution,
$\phi(v,J)$, of newly formed H$_2$ is strongly concentrated toward a
small number of states (see Black \& van Dishoeck [1987] for models with
pumping of specific levels), there will be little chance to detect signs
of it in the level distributions of H$_2$ in PDRs.

However, we should emphasize that the value of the H$_2$ formation rate
coefficient, $R_{gr}$, does strongly affect the total emission from a
PDR, since it governs its molecular abundance structure, and thereby the
fraction of the incident FUV energy that is reemitted by H$_2$, as
opposed to dust.

\begin{figure}[h]
\centering
\epsfig{file=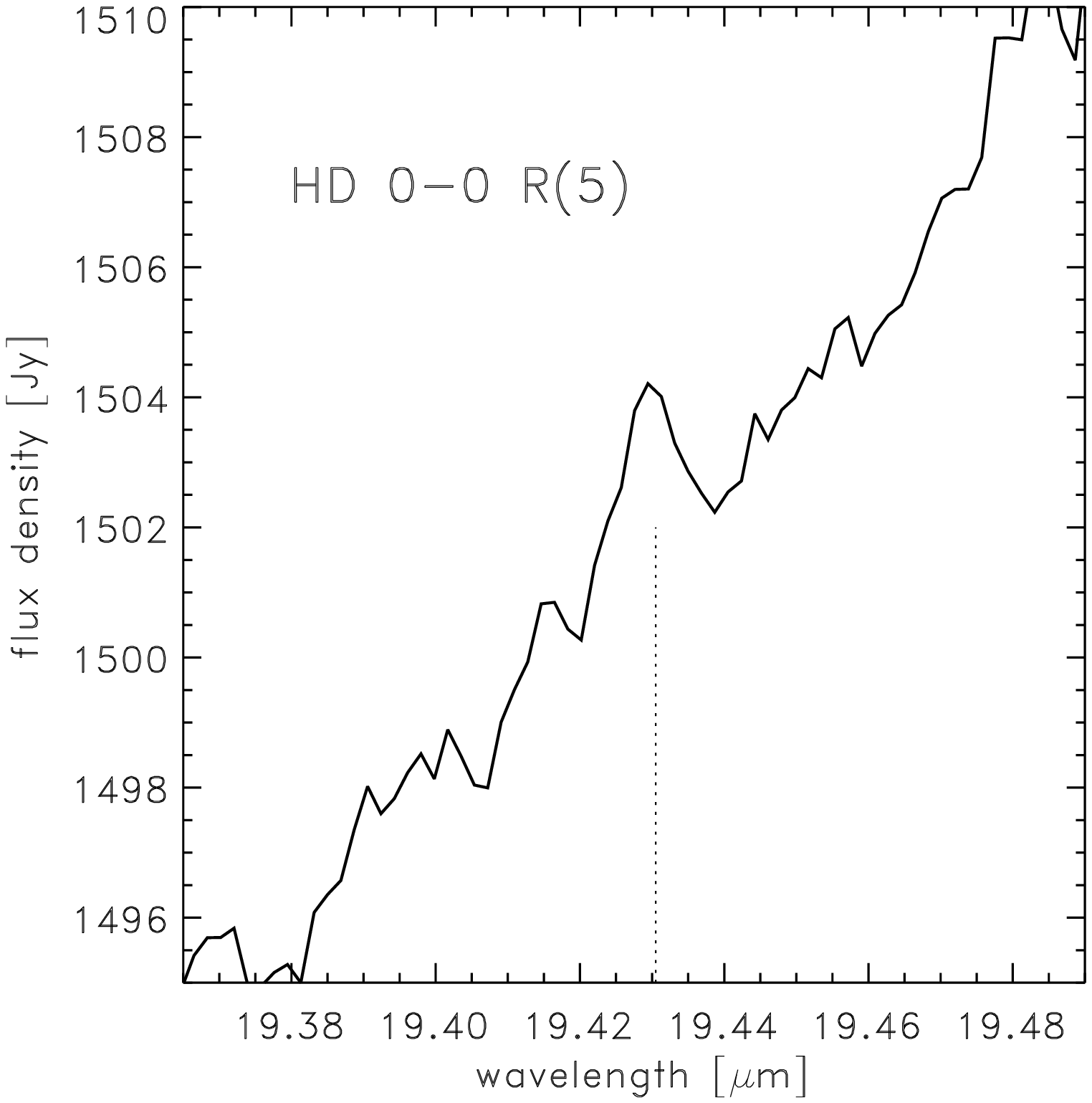,width=0.48\textwidth} 
\epsfig{file=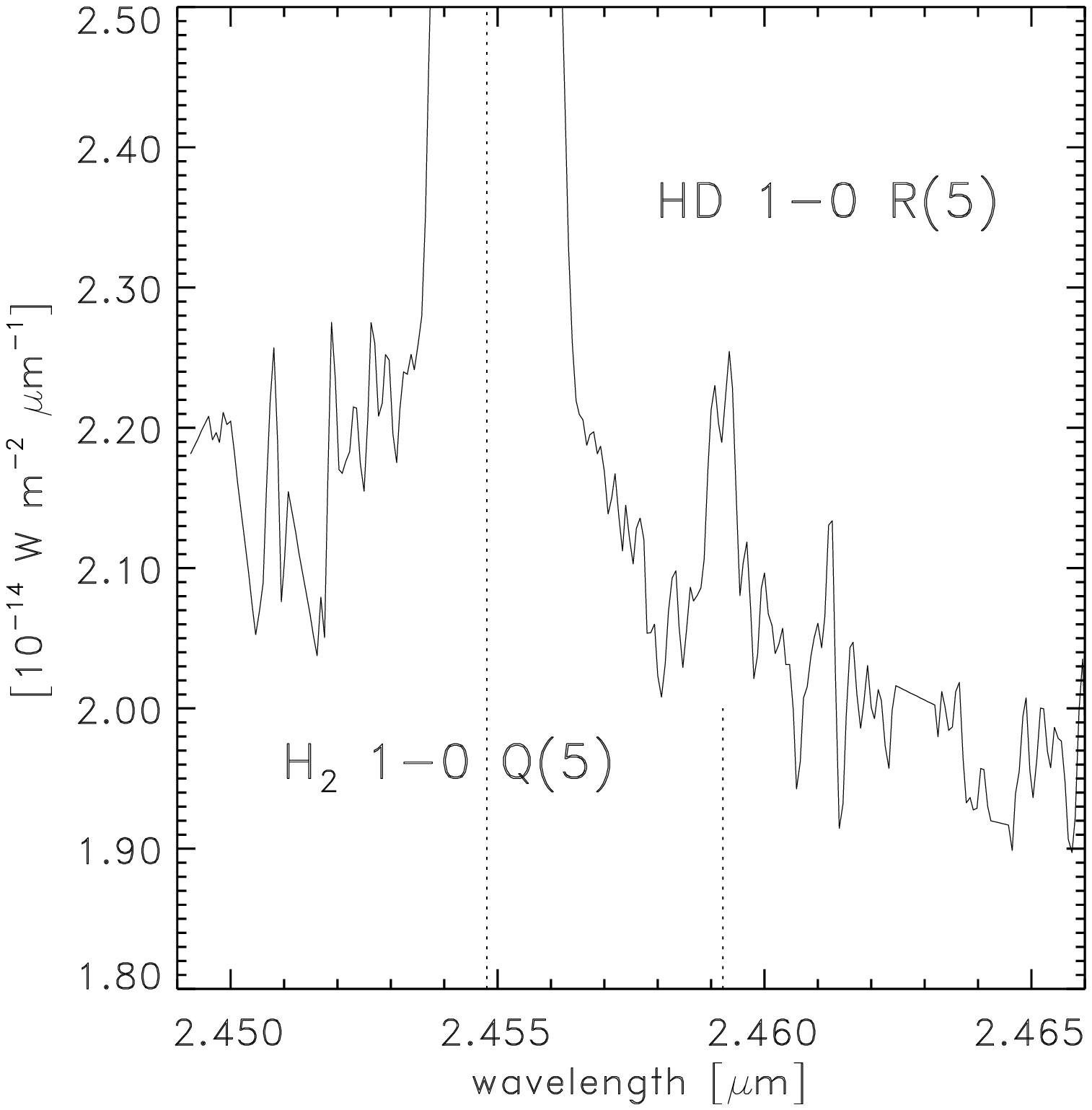,width=0.48\textwidth} 
\vspace{-.2cm}
\caption{
  Detection of interstellar HD emission,
  obtained with the ISO SWS ({\it left}) and the UKIRT CGS4 ({\it right}) toward
  Orion Peak 1.}
\label{fig:hd}
\end{figure}

\section{HD Emission from the Orion Outflow}

The ideal molecule for measuring the interstellar deuterium abundance is
HD, whose low rotational and ro-vibrational transition lines lie in the
near- and mid-infrared.  We used ISO and the United Kingdom Infrared
Telescope (UKIRT) to search for HD lines toward Peak 1.  With ISO we
there detected one faint HD line, 0--0 R(5) at 19.4 $\mu$m 
(Fig.\ 5, Bertoldi et al.\ 1999), and with UKIRT we subsequently
found the 1--0 R(5) line at 2.46 $\mu$m, providing the first observation
of HD emission with a ground-based telescope, and the first measurement
of the rotation-vibrational excitation of HD in the ISM (Ramsay Howat et
al., in prep.).

Surprisingly, we found that the HD excitation traced by the two observed
transitions exactly mirrors the excitation of H$_2$, despite the fact
that the much faster radiative transitions of HD place the critical
densities of its vibrationally excited states significantly above those of
H$_2$, and much higher than the estimated gas density in the outflow
region.  The cause for the similarity of the H$_2$ and HD excitation is
unresolved. It may imply a strong role of non-thermal excitation
mechanisms, such as the direct coupling of the HD excitation to that of
H$_2$ through the exchange reaction 
$\rm H_2$\,+\,D\,$\rightleftharpoons$\,HD\,+\,H.  
Current shock models are unable to explain the observed excitation of HD.

The similarity of the HD and the well-traced H$_2$ excitation permits
the yet most accurate determination of the [HD]/[H$_2$] abundance ratio
in the warm, shocked gas, and with a 30\% correction for chemical
depletion of HD relative to H$_2$ (due to an asymmetry in the
above-mentioned exchange reaction -- see Bertoldi et al.\ [1999], 
Ramsay Howat et al.\ in prep.) we are able to estimate a deuterium
abundance [D]/[H]\,=\,(5.1\,$\pm$\,1.5)\,$\times 10^{-6}$.

This value is low compared with previous D and H absorption measurements
towards nearby stars, which found values higher by a factor of two to
three.  But it is still consistent with two other measurement of the
deuterium abundance towards Orion (Wright et al.\  1999; Jenkins et al.\
1999). This indicates that in the Orion region, either stars have burnt
up a large fraction of the primordial deuterium, or more likely, that
there is significant trapping of deuterium on dust grains.  Future
observations from space and from the SOFIA airborne observatory will
eventually tell.


\acknowledgments

We are thankful to G.~Pineau des F\^orets, J.~Black, E.~van Dishoeck,
and C.~Wright for valuable input, and to the SWS Data Center at MPE,
especially to E.\ Wieprecht, for their support.  BTD was supported in
part by NSF grant AST-9619429.


\end{document}